\documentclass[a4paper]{mcs13}

\usepackage{tabularx} 
\usepackage{graphicx} 
\usepackage[intlimits]{amsmath}
\usepackage{amssymb} 
\usepackage{times}
\usepackage{soul} 
\usepackage[square,numbers,sort&compress]{natbib} 
\usepackage[labelfont=bf,font=normalsize]{caption}  
\usepackage{subcaption}
\usepackage{epstopdf} 
\usepackage{hyperref} 
\usepackage{siunitx}
\hypersetup{
    colorlinks=true,
    citecolor =blue,
    linkcolor=blue,
    urlcolor =blue
}

\usepackage{ulem}
\usepackage{pdfpages}

\begin{document}

\title{Direct Numerical Simulation of Hydrogen Combustion in a Real-Size IC Engine}

\authors{Bogdan A. Danciu*, George K. Giannakopoulos*, Mathis Bode** and Christos E. Frouzakis*}

\corrauthoremail{cfrouzakis@ethz.ch}

\address{*CAPS Laboratory, Department of Mechanical and Process Engineering,\\ETH Z\"{u}rich, 8092 Z\"{u}rich, Switzerland}
\address{**Jülich Supercomputing Centre, Forschungszentrum Jülich GmbH,\\ 52428 Jülich, Germany}

\begin{abstract} 

This study presents the first Direct Numerical Simulation (DNS) of hydrogen combustion in a real-size internal combustion engine, investigating the complex dynamics of ignition, flame propagation, and flame-wall interaction under engine-relevant conditions. The simulation focuses on ultra-lean hydrogen operation at equivalence ratio $\phi=0.4$ and 800 rpm, utilizing a state-of-the-art spectral element solver optimized for GPU architectures. The computational domain encompasses the full engine geometry.
Results highlight the strong coupling between the flame dynamics and the coherent flow structures during early flame kernel development, while differential diffusion effects lead to increased reactivity at positive flame curvatures, a phenomenon that has only been studied in canonical configurations of freely propagating hydrogen/air flames. As the flame approaches the walls, distinct behavior is observed during head-on and side-wall quenching scenarios, characterized by different spatial distributions of wall heat flux. The findings provide insights into hydrogen combustion in real engines, essential for the development of clean and efficient hydrogen-fueled powertrains.

\end{abstract}

\section{Introduction}

The imperative to decarbonize the global energy sector has intensified interest in hydrogen as a carbon-free energy carrier for internal combustion engines (ICEs) \cite{Verhelst2009, Wrbel2022}. The unique properties of hydrogen, including the capacity for renewable production, high energy density, and broad flammability limits position it as a compelling alternative to fossil fuels, particularly for heavy-duty transportation and industrial applications where battery electrification faces significant challenges~\cite{Stpie2021}. The adoption of hydrogen ICEs offers a promising pathway toward carbon neutrality while leveraging existing manufacturing infrastructure and engineering expertise~\cite{Dreizler2021}. 
Hydrogen's wide flammability limits and high flame speeds enable ultra-lean combustion strategies in ICEs, simultaneously reducing nitrogen oxide (NOx) emissions and improving thermal efficiency. NOx emissions decrease substantially when operating at fuel-air equivalence ratios ($\phi$) below 0.5~\cite{DAS1991, RamalhoLeite2023}, though significant research challenges remain in understanding early flame kernel development, turbulent flame propagation dynamics, and flame-wall interactions in real-size ICEs. 

Understanding the fundamental processes of hydrogen combustion is critical to address these challenges. One of the most crucial stages is the early flame kernel development, which determines the success of ignition and subsequent flame propagation~\cite{Chu2023}. The dynamics of this phase are influenced by local turbulence, chemical kinetics, and the interaction of the flame with the surrounding flow field. In ICEs, this interaction is particularly important as the flow is dominated by coherent large-scale structures~\cite{Hill1994, Bore2014} that significantly impact flame kernel development. Furthermore, as the flame propagates, its interaction with the engine walls plays a significant role in determining thermal losses and pollutant formation. As the flame encounters the combustion chamber surfaces, these interactions intensify, leading to reduced reactivity near the walls and eventual flame quenching, processes that critically impact overall engine efficiency and durability.

The experimental characterization of hydrogen flames has evolved from early fundamental studies in pressure chambers, where Schlieren imaging provided insights into laminar flame dynamics~\cite{Bauwens2017, Xie2022}, to increasingly complex engine-relevant conditions with a focus on ultra-lean combustion strategies. In-cylinder measurements, while challenging, have provided valuable insight into the behavior of hydrogen combustion, particularly at equivalence ratios below 0.5, where NOx emissions are significantly reduced. Optical diagnostics have shown how the large-scale flow structures, such as the tumbling motion, significantly influence flame kernel convection and distortion~\cite{Salazar2011}. The transition from conventional hydrocarbon fuels to hydrogen was investigated using UV-based flame luminosity measurements, which highlighted the distinctive characteristics of hydrogen, namely increased flame front wrinkling and accelerated propagation speed, which is particularly important for understanding cycle-to-cycle variations (CCV) in early flame development. While OH-LIF techniques have traditionally served as a reliable tool for visualizing flame structure, their effectiveness diminishes significantly in lean hydrogen flames due to lower signal intensity, which is particularly challenging for $\phi < 0.5$ . This limitation has spurred the development of alternative diagnostic approaches, with SO$_2$-LIF emerging as a promising technique for resolving flame fronts under engine-relevant conditions at elevated pressures and temperatures~\cite{Honza2017, Welch2024}.

Complementing experimental efforts, computational studies have provided additional insights into hydrogen combustion. 
Direct Numerical Simulation (DNS) studies, while largely limited to canonical configurations due to computational cost, have revealed fundamental aspects of hydrogen flame-turbulence interactions~\cite{Falkenstein2020}, differential diffusion effects~\cite{Chu2023}, and the complex interplay between turbulent burning velocity and flame structure at varying Karlovitz numbers~\cite{Lee2022}. However, DNS studies of hydrogen combustion have not yet been extended to real-size engine geometries. The present study addresses this knowledge gap by performing the first ever DNS of hydrogen combustion in a real-size ICE, enabled by recent advances in high-performance computing and the use of GPU-accelerated supercomputers. Building upon the framework established for non-reactive engine simulations~\cite{Giannakopoulos2022, Danciu2024}, this work investigates the ignition and flame propagation in a slightly modified geometry of the optically-accessible TU Darmstadt (TUDa) engine~\cite{Baum2014} operating with ultra-lean hydrogen mixtures at partial load conditions. Utilizing a state-of-the-art spectral element solver optimized for GPU architectures~\cite{nekCRF}, the simulation resolves the relevant scales, providing insights into the coupled dynamics of flame kernel development, turbulent flame propagation, and flame-wall interaction under engine-relevant conditions.

 The rest of the paper is organized as follows. Section~\nameref{sec:methodology} describes the computational methodology including the engine geometry and DNS setup.
The subsequent sections discuss the \nameref{sec:ign}, \nameref{sec:propagation} and \nameref{sec:fwi}. The main conclusions and future research directions are outlined in ~\nameref{sec:conclusions}.

\section{Numerical methodology} \label{sec:methodology}

The TUDa engine is a single cylinder spark ignition engine with a pent roof, four-valve head and an inlet duct designed to promote tumble flow generation. The cylinder of the square engine has a bore of $B=86$~mm, and the engine was designed to provide well-defined boundary conditions and reproducible operation \cite{Welch2024}. Optical access is provided by the quartz glass liner and flat piston. Extensive experimental campaigns have analyzed in depth the flow characteristics under motored operation at 800, 1500 and 2500 rpm at partial and full load conditions (intake pressure $p_{in}=0.4$ and 0.95~bar, respectively) (e.g. ~\cite{Renaud2018,Schmidt2023}.)
Recent experiments of fired operation with lean hydrogen \cite{Welch2024} focused on the role of thermo-diffusive and hydrodynamic instabilities and the effect of equivalence ratio on the early flame development at 800~rpm and full or partial load. 

Expanding on the foundation of \cite{Giannakopoulos2022}, the focus of this paper is the fired operation of case H$_2$-B in \cite{Welch2024}, i.e. partial load with hydrogen at equivalence ratio $\phi=0.4237$. Previous DNS of motored operation in the TUDa engine \cite{Giannakopoulos2022,Danciu2024} employed fields at inlet valve closure that were generated by experimentally-validated multi-cycle LES as initial conditions. A similar approach is used in this work, where due to the lack of precursor reactive LES, the flow and temperature fields are obtained by interpolating the motored engine DNS of ~\cite{Giannakopoulos2022} at 660~CAD. The initial gas composition is considered uniform, consisting of a hydrogen-air mixture with equivalence ratio $\phi=0.4237$. The no-slip inert walls are assumed to be isothermal at $T_w=333.15$~K, except for the spark plug electrodes which are considered adiabatic. The kinetics are described by the detailed mechanism of Burke et al.~\cite{Burke2011}, consisting of 9 species in 23 reactions, while diffusive processes are modeled using the mixture-averaged formulation without considering the Soret effect. The simulation covered the range between 660 and 700.4~CAD.

\begin{figure}[h!]
    \centering
    \includegraphics[width=0.9\linewidth]{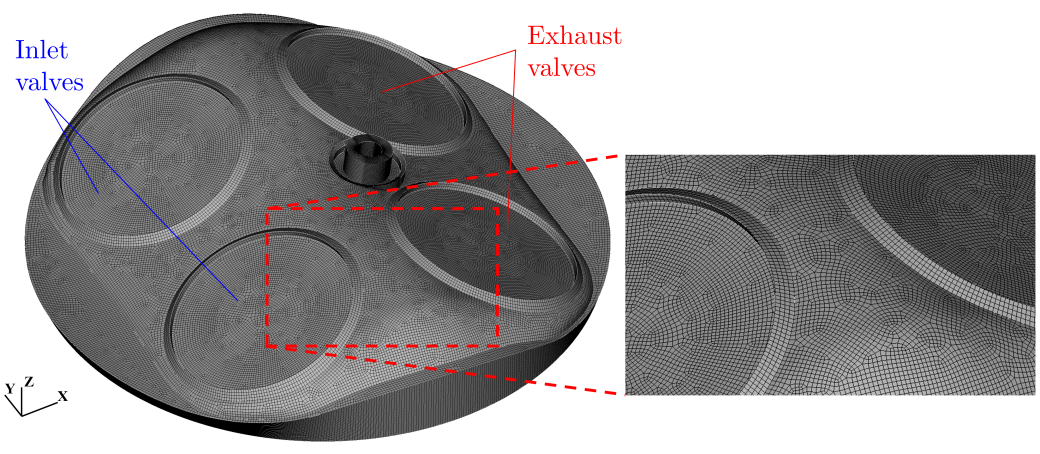}
    \caption{ Engine head mesh with the spectral element skeleton (i.e. without the points within the elements) of the conformal hexahedral mesh.}
    \label{fig:hex_mesh}
\end{figure}

To simplify mesh generation and relax excessive time step restrictions related to CFL, the engine geometry was modified by removing the long crevice volume and the injector cap. 
The grid generation approach was revised, moving from a tetrahedral-to-hexahedral conversion used in the non-reactive simulations to a purely hexahedral mesh generated using Coreform Cubit 2024.8, significantly improving the overall mesh quality. Figure~\ref{fig:hex_mesh} shows the skeletal mesh (i.e. without the points within each spectral element) of the computational domain at 660 CAD, highlighting the locally structured nature of the hexahedral conforming mesh. The mesh on the lower surface of the head was extruded to the piston to create tensor product element layers that can accommodate the vertical mesh deformation due to piston motion while minimizing element distortion. 
At the piston six refinement layers of elements were used, resulting in a $\SI{1.55}{\micro\metre}$ wall-normal distance of the first point. 

\begin{figure}[h!]
    \centering
    \begin{subfigure}{0.625\textwidth}
        \includegraphics[width=\textwidth]{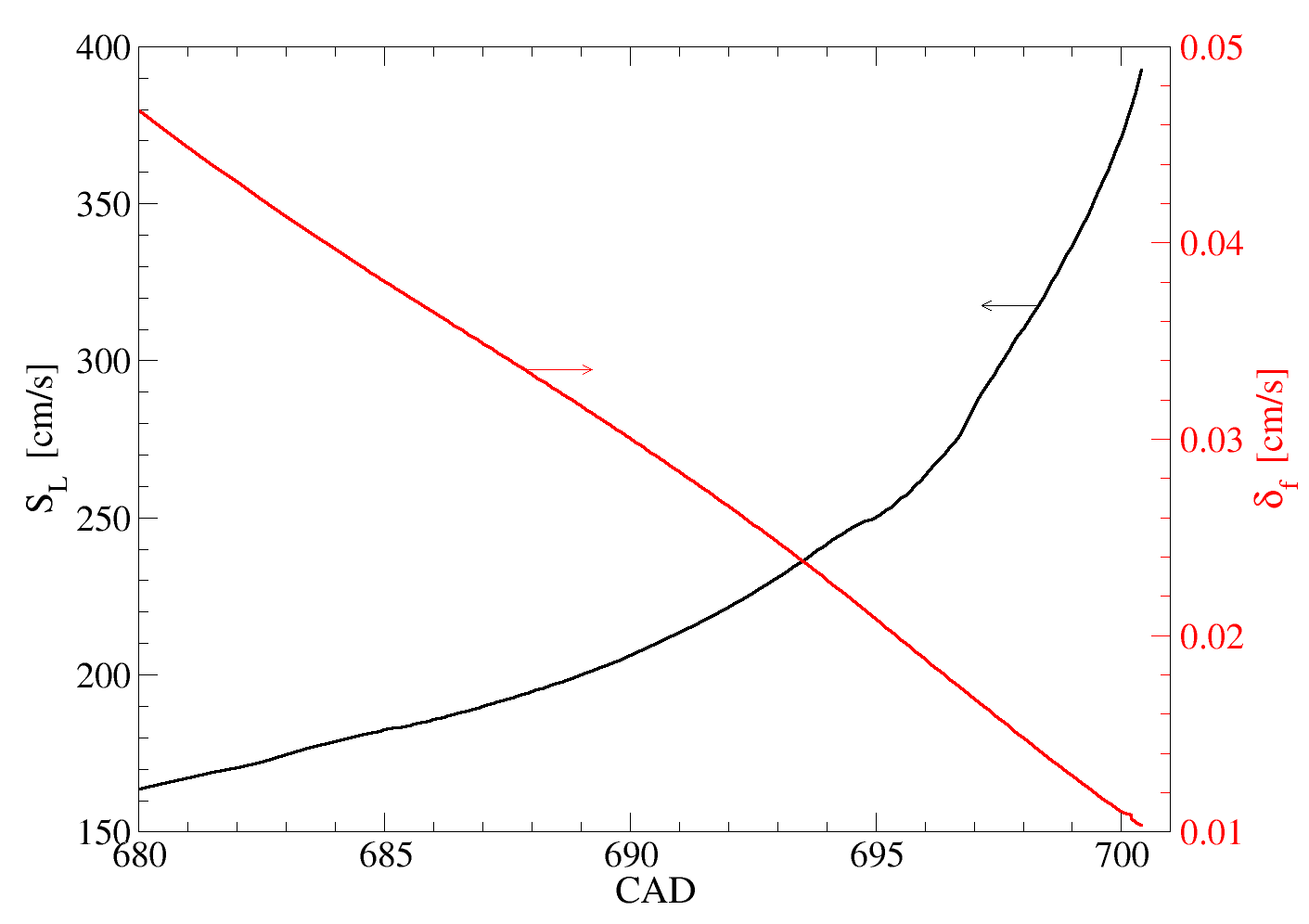}
        \caption{}
        \label{fig:df_SL}
    \end{subfigure}
    \begin{subfigure}{0.675\textwidth}
        \includegraphics[width=\textwidth]{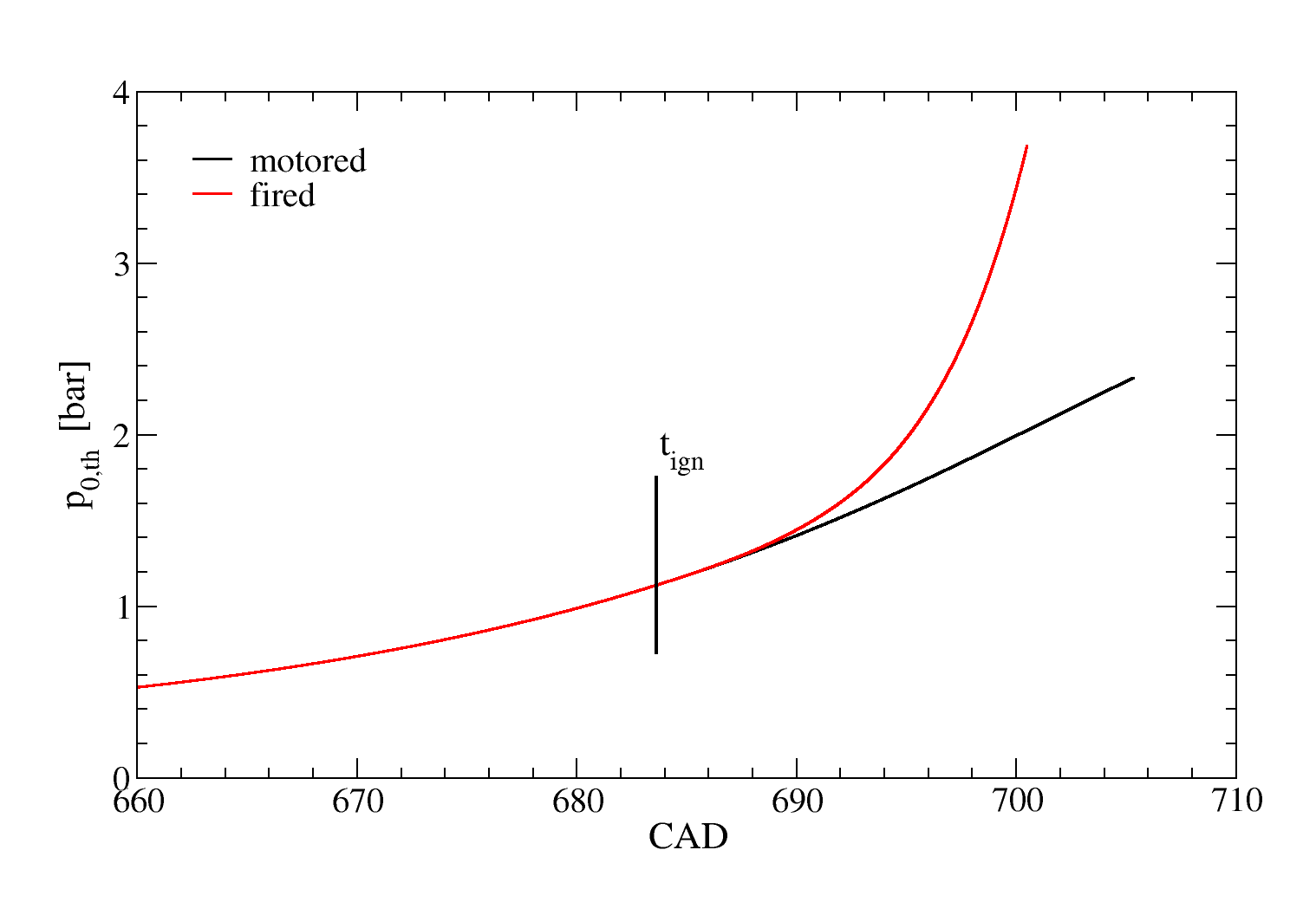}
        \caption{}
        \label{fig:p0th}  
    \end{subfigure}
    \caption{Evolution of (a) laminar flame thickness $\delta_f$ and flame speed $S_L$ computed at the instantaneous thermodynamic conditions and (b) thermodynamic pressure during motored and fired operation.    }
    \label{fig:df_SL_p0th}
\end{figure}

Since the mesh is nonuniform and the flame orientation changes during propagation, it is not straightforward to assess how well the flame structure is resolved.
The boundary layer is fully resolved and flame-wall interaction is comprehensively captured. Additionally, the mesh is refined around the spark plug to accurately capture the resolution-demanding phase of flame kernel formation. 
At spark time, the laminar flame thickness is around \SI{410}{\micro\metre} and decreases to \SI{101}{\micro\metre} at 700 CAD due to compression (Fig.~\ref{fig:df_SL_p0th}a). Both flame speed and thickness are computed at the instantaneous thermodynamic pressure and unburned mixture temperature, which evolve differently between motored and fired conditions (Fig.~\ref{fig:df_SL_p0th}b). With 8 points within the flame thickness, the grid resolution is considered adequate to resolve both flow and flame dynamics at least up to 698~CAD.
The final mesh consists of $E=7.4$ million spectral elements, within each of which the solution is approximated using order $N=7$ Legendre polynomials through the Gauss--Lobatto--Legendre quadrature points, resulting in 2.5 billion unique discretization points.

The simulations were performed with the low Mach number reactive flow solver nekCRF \cite{nekCRF}, developed as an extension to the spectral element CFD solver nekRS \cite{nekRS}. nekCRF can run on CPUs as well as GPU-accelerated hardware that support CUDA, HIP, or OpenCL through the OCCA library unified API \cite{occa}. The geometric flexibility of the spectral element method is essential for the accurate representation of the complex time-varying geometry of the TUDa engine. nekCRF uses the time splitting scheme of \cite{TLO97}, where the thermochemistry subsystem (species and energy equations) is integrated implicitly as a fully-coupled system using CVODE \cite{sundials}, while the hydrodynamic subsystem (continuity and momentum) is advanced using a third-order backward differentiation formula to approximate the time derivative. Additionally, nekCRF employs efficient fuel-specific kernels for the computation of the chemical source terms, thermodynamic properties, and transport coefficients generated by KinetiX \cite{kinetix}. The simulation used 150 nodes of the JUWELS Booster at the J\"{u}lich Supercomputing Centre, each equipped with 2 AMD EPYC Rome CPUs with 48 cores and 4 NVIDIA A100 GPUs with 40~GB memory, requiring 12,250 node hours in total. Visualization and data analysis were performed using VisIt \cite{visit_2012}.

Except otherwise noted, the results are reported in normalized with respect to the bore $L_{ref}=B=\SI{86}{\milli\metre}$, the maximum piston speed $U_{ref}=V_{p,max}=\SI{3.752}{\metre/s}$, the wall temperature $T_{ref}=\SI{333.15}{K}$, and the properties of the unburned mixture.

\section{Ignition} \label{sec:ign}

Ignition at the spark plug is modeled through a spatio-temporally varying energy deposition following the approach described in \cite{Falkenstein2020}. The spark source $\dot{S}_{ign}$ is initiated at $t_{ign}=683.4$~CAD bTDC as in the experiments \cite{Welch2024}, has total ignition energy $E_{0}=\SI{0.1}{\milli\joule}$, kernel radius $r_{ign}=\SI{1.4}{\milli\metre}$ and duration $\tau_{ign}=\SI{0.2}{\milli\second}$ (4.5 CAD), and is distributed radially according to 
\begin{equation}
\dot{S}_{ign}(\hat{t}, r) = 
\begin{cases}
\frac{\psi(\hat{t})}{\int_0^1 \psi d\hat{t}} \frac{E_0}{c_p \tau_{ign} V_{ign}} \left[1 - \exp\left(-\left[\frac{\pi}{2} - \frac{r}{r_{ign}}\right]^4\right)\right], & r \leq \frac{\pi}{2}r_{ign} \\ \\
0, & \text{otherwise}
\end{cases}
\end{equation}
where $\hat{t} = (t - t_{ign})/\tau_{ign}$ is the non-dimensional time, $V_{ign} = 4/3\pi r_{ign}^3$ the reference volume, and $\psi(\hat{t}) = \frac{2\hat{t}^2}{C^2}\frac{\ln(\hat{t})}{\ln(C)}$ the time-dependent scaling factor with $C = 0.603$. The smooth temporal and spatial distribution of the ignition energy respect the low-Mach number restriction.

\begin{figure}[h!]
    \centering
    \begin{subfigure}{0.625\textwidth}
        \includegraphics[width=\textwidth]{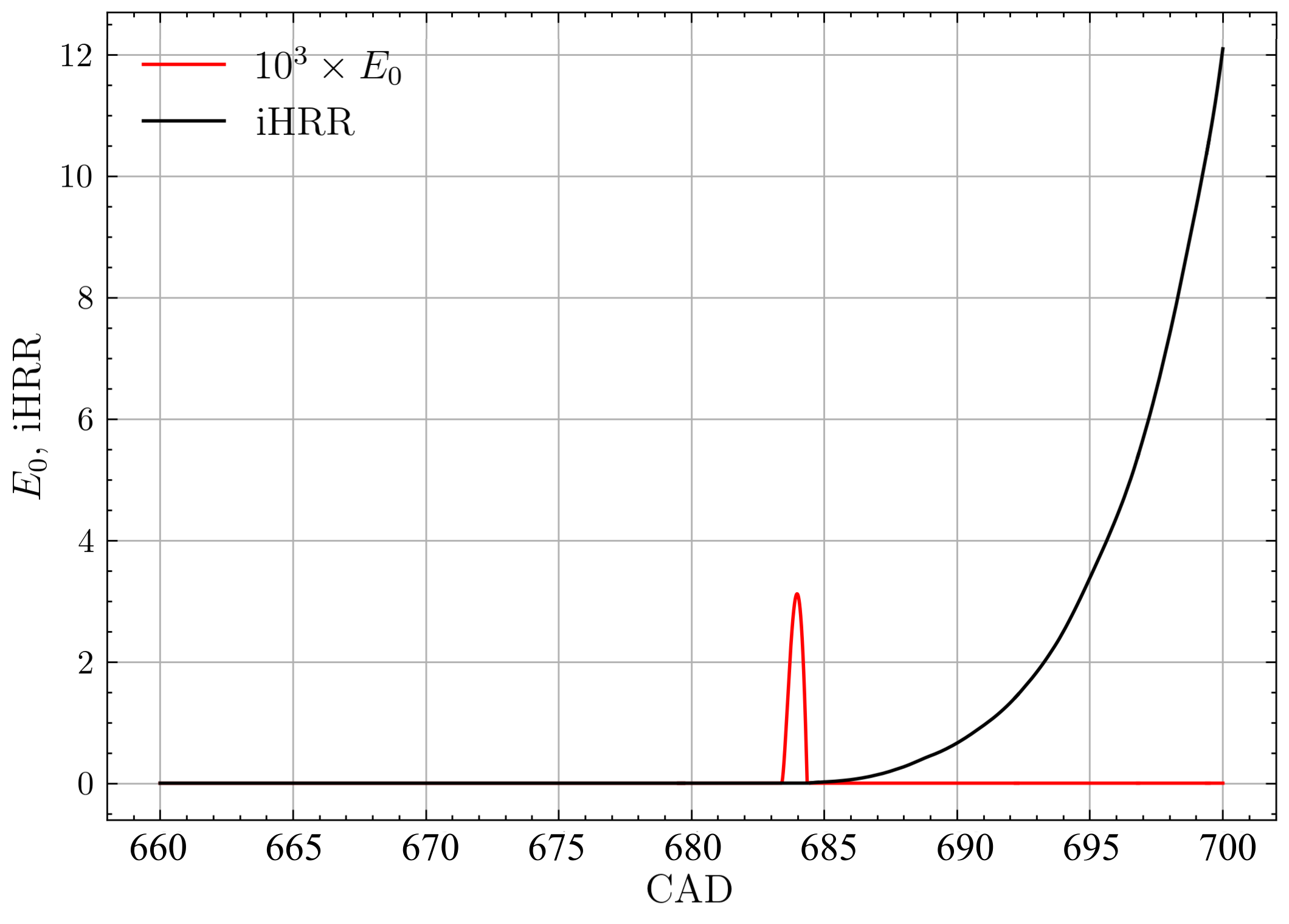}
        \caption{}
        \label{fig:hrr}
    \end{subfigure}
    \begin{subfigure}{0.605\textwidth}
        \includegraphics[width=\textwidth]{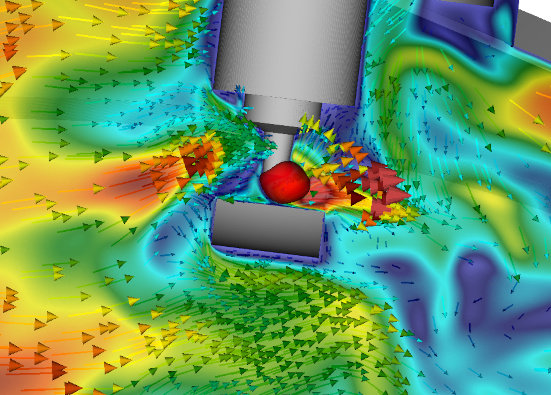}
        \caption{}
        \label{fig:kernel}
    \end{subfigure}    
    \caption{(a) Temporal evolution of the normalized integrated heat release rate (iHRR) and cumulative energy deposition ($E_0 \times 10^3$), (b) flame kernel defined by the $Y_{H_2}=0.0048$ iso-surface superimposed on the in-plane velocity magnitude and vectors along the tumble plane at 1.1 CAD aSOI.}
    \label{fig:hrr_kernel}
\end{figure}

The spark gap is located in a region where the flow is affected by the tumble motion, with strong clockwise velocity components as can be seen in Fig.~\ref{fig:hrr_kernel}(b). The energy deposition leads to a rapid rise in the local temperature, generating a sufficient pool of radicals for the initiation of a self-sustaining flame kernel. The transition is marked by the onset of exponential growth in the integral heat release rate (iHRR, Fig.~\ref{fig:hrr_kernel}a) , leading to the faster increase of the thermodynamic pressure compared to motored operation (Fig.~\ref{fig:df_SL_p0th}b). 

In the following, the flame surface is defined by the $Y_{H_2}=0.0048$ iso-surface, which corresponds to $C=1-Y_{H_2}/{Y_{H_2,u}}=0.6$ when expressed in terms of the progress variable $C$ corresponding to the maximum rate of heat release for this mixture. The flame surface at 1.1 CAD after start of ignition (aSOI) is shown in Fig.~\ref{fig:hrr_kernel}(b), superimposed on the velocity magnitude distribution and vectors on the tumble plane. The small kernel is initially distorted by the spark electrodes and wrinkled by the local turbulence. The orientation of the spark plug is expected to play an important role in the formation and shape of the initial flame kernel, as the ground electrodes might obstruct or expose it to the strong convective flow.

\section{Flame propagation} \label{sec:propagation}

\subsection{Phenomenology}

\begin{figure}[h!]
    \centering
    \includegraphics[width=\linewidth]{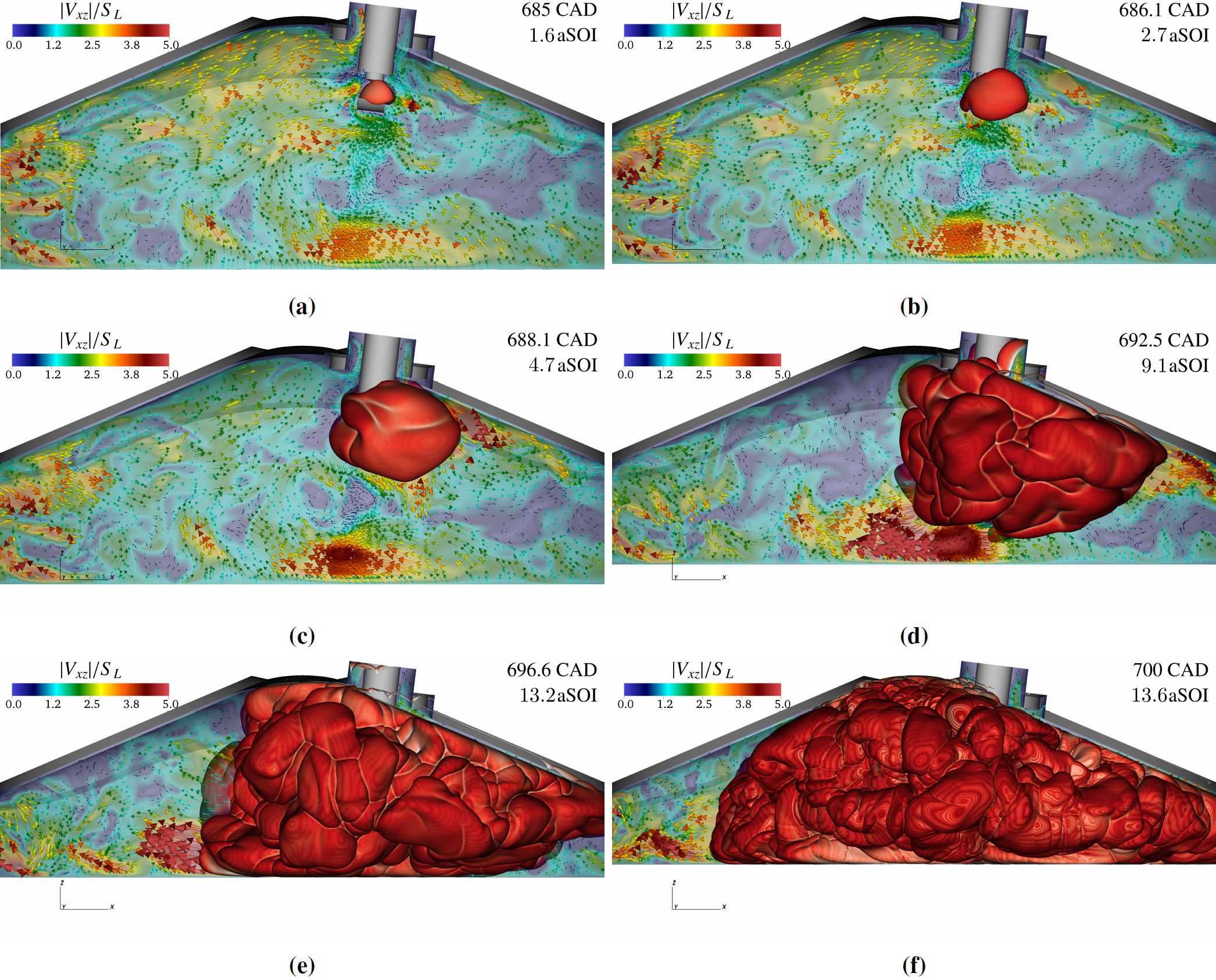}
    \caption{Evolution of the flame kernel defined by the $Y_{H_2}=0.0048$ iso-surface superimposed on the velocity magnitude distribution on the $y=0$ plane normalized by the laminar flame speed $S_L$ at the instantaneous thermodynamic conditions at (a) 1.6, (b) 2.7, (c) 4.7, (d) 9.1, (e) 13.2, (f) 13.6 CAD aSOI. The flame kernel is colored by the heat release rate.}
    \label{fig:flame_prop}
\end{figure}

The flame evolution is strongly influenced by the local flow field and geometry. Initially, the flame kernel develops in a region dominated by the tumble motion, which generates a strong flow from left to right as shown by the velocity vectors in Figs.~\ref{fig:hrr_kernel}(b) and \ref{fig:flame_prop}(a). Despite thermal expansion effects, the flame front is pushed towards the exhaust valves since the gas expansion cannot overcome the incoming flow. 
As compression progresses, the large coherent tumble vortex breaks down into smaller scale turbulent structures, redistributing the kinetic energy towards a more isotropic state \cite{Giannakopoulos2019, Giannakopoulos2022}, causing a deceleration of the flow upstream of the flame. After 9.1 CAD aSOI (Fig.~\ref{fig:flame_prop}d), the flame thermal expansion is strong enough to balance and locally reverse the flow direction. Beyond this point, the flame starts propagating in all directions at an increasing rate, while being wrinkled by the continuously intensifying turbulent flow it encounters, both due to compression and the ongoing tumble breakdown process.

As the pressure in the combustion chamber increases, the flame morphology is also affected by differential diffusion effects, which are expressed in the form of cellular structures on the flame front. These structures are reminiscent of thermo-diffusive instabilities, demonstrating an increase (decrease) in the heat release rate for positive (negative) flame curvature, and known to be particularly pronounced in lean hydrogen flames. Finally, when the flame reaches the piston surface around 695 CAD (11.6 CAD aSOI), the local flow patterns and the confined geometry strongly affect the flame shape, leading to head-on and side-wall quenching at various locations within the combustion chamber, phenomena that will be discussed in more detail in the following.

\subsection{Flame speed}

We now examine the flame propagation characteristics in more detail, focusing on the evolution and the interplay between the effects of reaction, diffusion and flow-flame interactions. The displacement speed is a quantity that provides valuable insights into the coupled effects of reaction, diffusion, and flow-flame interactions. It is defined as the speed of a scalar iso-surface (here $Y_\mathrm{H2}=0.0048$) relative to the flow of fresh reactants and can be computed directly using the right-hand side of the transport equation \cite{Giannakopoulos2015}
\begin{equation}
    S_d = \frac{1}{|\nabla Y_{H_2}|} \frac{DY_{H_2}}{Dt} \Bigg |_{Y_{H_2}=0.0048}.
\end{equation}
It can be split into contributions from the normal flame propagation $S_{d,rn}$ and tangential diffusion $S_{d,t}$ as 
\begin{align}
     S_{d,rn} &= \frac{1}{\rho |\nabla Y_{H_2}|} \left[ \dot{\omega}_{H_2} 
- \mathbf{n}\cdot \nabla \left( \rho D_{H_2} |\nabla Y_{H_2}|\right)\right], \\[1em]
    S_{d,t} &= -D_{H_2} \kappa \label{eq:sdt},
\end{align}
where $\rho$ is the density, $\dot{\omega}_{H_2}$ the fuel consumption rate, $D_{H_2}$ the mixture-averaged diffusivity of $H_2$ and $\kappa = \nabla \cdot {\textbf{n}}$ the flame curvature with ${\textbf{n}}=Y_{H_2} / |\nabla Y_{H_2}|$ the flame normal vector pointing towards the unburned mixture. Positive (negative) curvature corresponds to convex (concave) segments of the flame front towards the unburned gas. 
Of particular interest is how these quantities evolve as the flame encounters increasingly intense turbulence due to compression and tumble breakdown, while simultaneously being affected by the thermo-diffusive imbalance effects and the time-varying thermodynamic conditions.

\begin{figure}[h!]
    \centering
    \includegraphics[width=\linewidth]{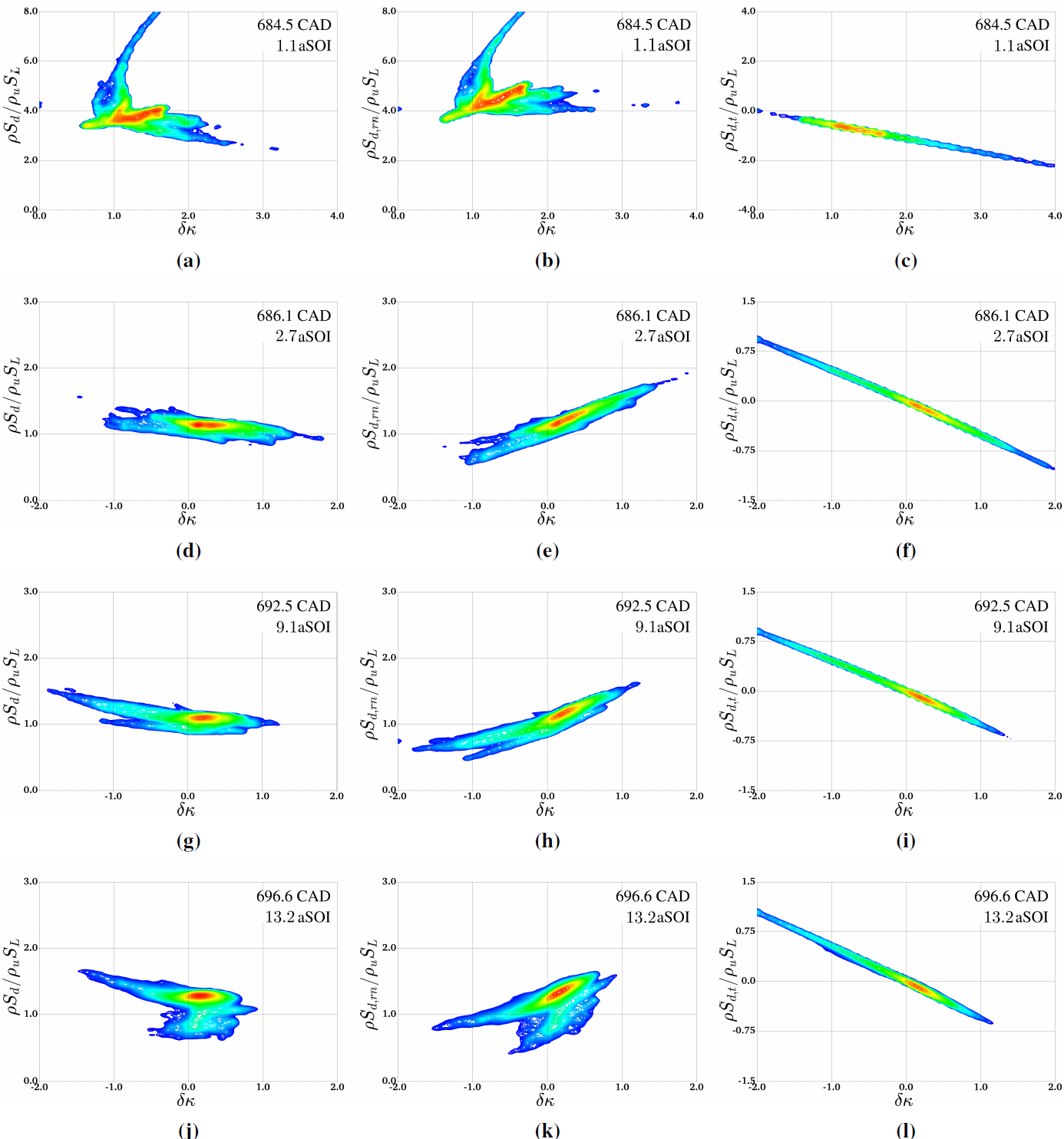}
    \caption{Joint pdfs of the density-weighted $S_{d}$, $S_{d,rn}$, $S_{d,t}$  as a function of curvature at (a-c) 1.1, (d-f) 2.7, (g-i) 9.1, (j-l) 13.2 CAD aSOI. The quantities are normalized by the unburned mixture density $\rho_u$, laminar flame speed $S_L$ and thickness $\delta$ of the laminar flame at the corresponding instantaneous thermodynamic conditions.}
    \label{fig:joint_pdfs}
\end{figure}

Figure~\ref{fig:joint_pdfs} presents the joint probability density functions (pdfs) of the density-weighted $S_{d}$, $S_{d,rn}$, $S_{d,t}$ as a function of curvature at various time instances. The quantities are normalized by the unburned mixture density $\rho_u$, laminar flame speed $S_L$ and thickness of the 1-D laminar flame $\delta$ at the instantaneous thermodynamic conditions (see Fig.~\ref{fig:df_SL_p0th}), such that their values can be directly compared. Early-on the flame kernel is largely affected by the energy deposition from the ignition source, and the flame speed acquires very high values, mostly driven by the $S_{d,rn}$ component (Figs.~\ref{fig:joint_pdfs}a,b). As the flame propagates and moves away from the ignition spot, the flame displacement speed becomes comparable to the laminar flame speed and the dependence on curvature becomes progressively more linear. The correlation between $S_{d,t}$ and curvature is linear and negative, as expected from the definition (\ref{eq:sdt}) and the low Lewis number mixture considered here. In contrast, the normal propagation component $S_{d,rn}$ exhibits a positive correlation with curvature, especially for positive curvatures, due to diffusion effects (e.g. Fig.~\ref{fig:joint_pdfs}e,h). The net effect of these competing mechanisms is the increased flame speed for negative curvatures (dominated by the high $S_{d,t}$ values), whereas for positive curvatures the increase in reactivity is balanced out by the negative tangential diffusion contribution, leading to flame speeds that are rather insensitive to curvature (Fig.~\ref{fig:joint_pdfs}g). These observations are in agreement with previous studies of freely propagating flames for the same mixture at engine-relevant conditions \cite{Chu2023}. When significant part of the flame surface reaches the walls, a branch with reduced reactivity and flame speed emerges (Fig.~\ref{fig:joint_pdfs}j,k), marking the phase where the flame is affected by interaction with the cold piston.

\section{Flame-wall interaction} \label{sec:fwi}

\subsection{Wall heat flux}

\begin{figure}[h!]
    \centering
    \includegraphics[width=0.9\linewidth]{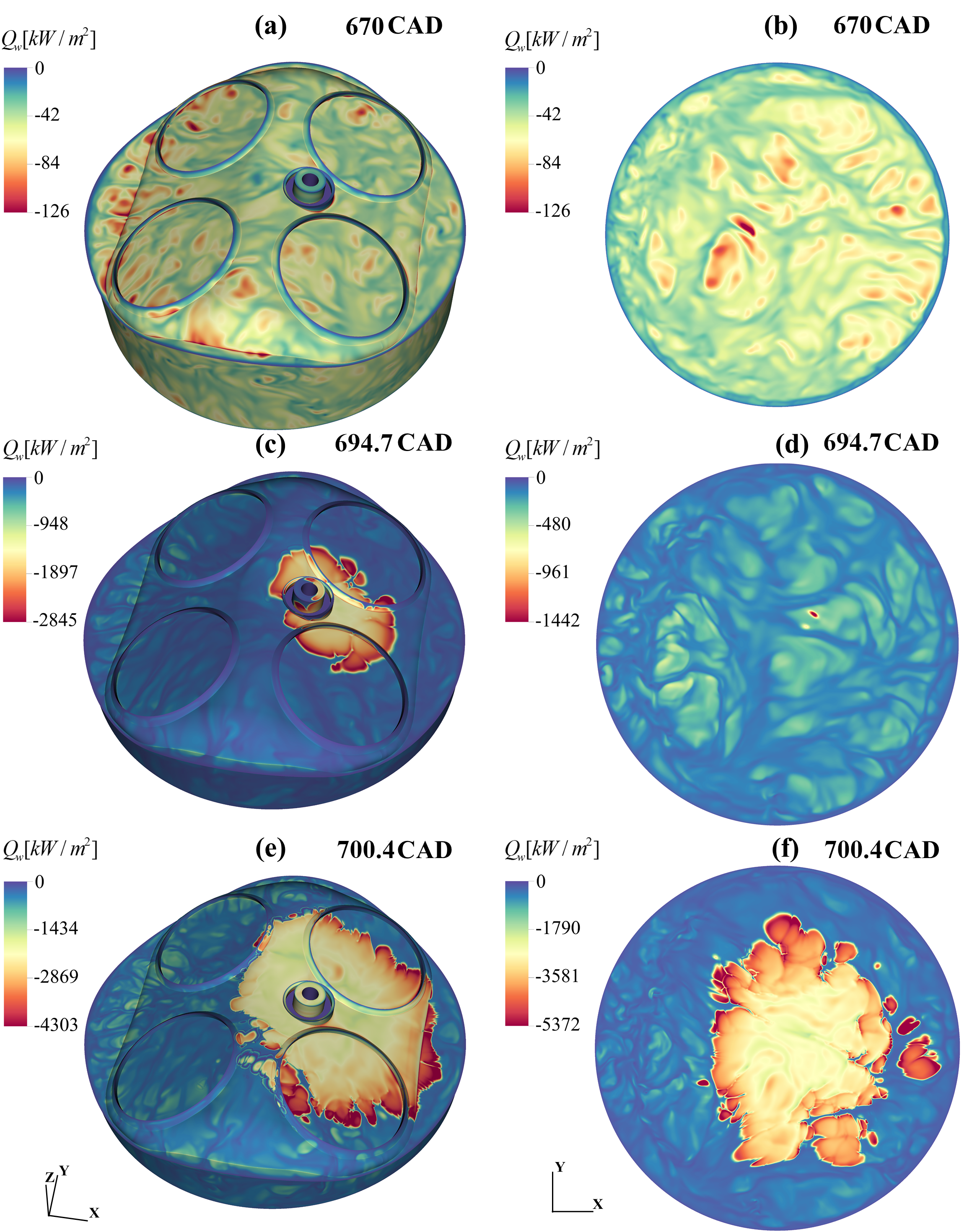}
    \caption{Iso-contours of heat flux along the cylinder walls at 670 CAD (top), 694.7 CAD (middle) and 700.4 CAD (bottom). Left: bird's eye view, right: piston view.}
    \label{fig:whf_walls}
\end{figure}

Figure~\ref{fig:whf_walls} depicts the iso-contours of the local wall heat flux on the cylinder walls and piston at different times during the compression and combustion phases. During mid compression at 670 CAD (Fig.~\ref{fig:whf_walls}a,b), the heat flux values are relatively low (up to a few hundred $kW/m^2$) and its spatial distribution is mainly determined by the in-cylinder flow structures. Similarly to what was observed during non-reactive simulations \cite{Giannakopoulos2022, Danciu2024}, the wall heat flux is higher where the flow is impinging/stagnating due to the large-scale tumbling motion, such as on the left side of the engine head (Fig.~\ref{fig:whf_walls}a) and the right side of the piston (Fig.~\ref{fig:whf_walls}b). When the flame first approaches the piston (Fig.~\ref{fig:whf_walls}d), a peak heat flux is observed that is about an order of magnitude larger (the change in the scale of the color map in each graph should be noted). Focusing on the shape of the heat flux profile at the point of impingement, it can be seen that the heat flux values are higher in the center of the contact area and lower around it. This indicates that the flame is approaching the wall perpendicularly in a head-on quenching (HOQ) mode. As the flame grows, the contact area of the flame with the walls of the combustion chamber increases, increasing the total amount of heat loss (i.e. the integrated local heat flux along the chamber walls). After the initial HOQ, the flame slides along the wall as it propagates and consumes the fresh mixture during the side-wall quenching (SWQ) phase (Fig.~\ref{fig:whf_walls}c,e,f). The spatial distribution of the local heat flux is now different to the head-on collision: the peak values are located along the edge surrounding the contact region and not at its center.

\begin{figure}[h!]
\centering
\includegraphics[width=0.575\linewidth]{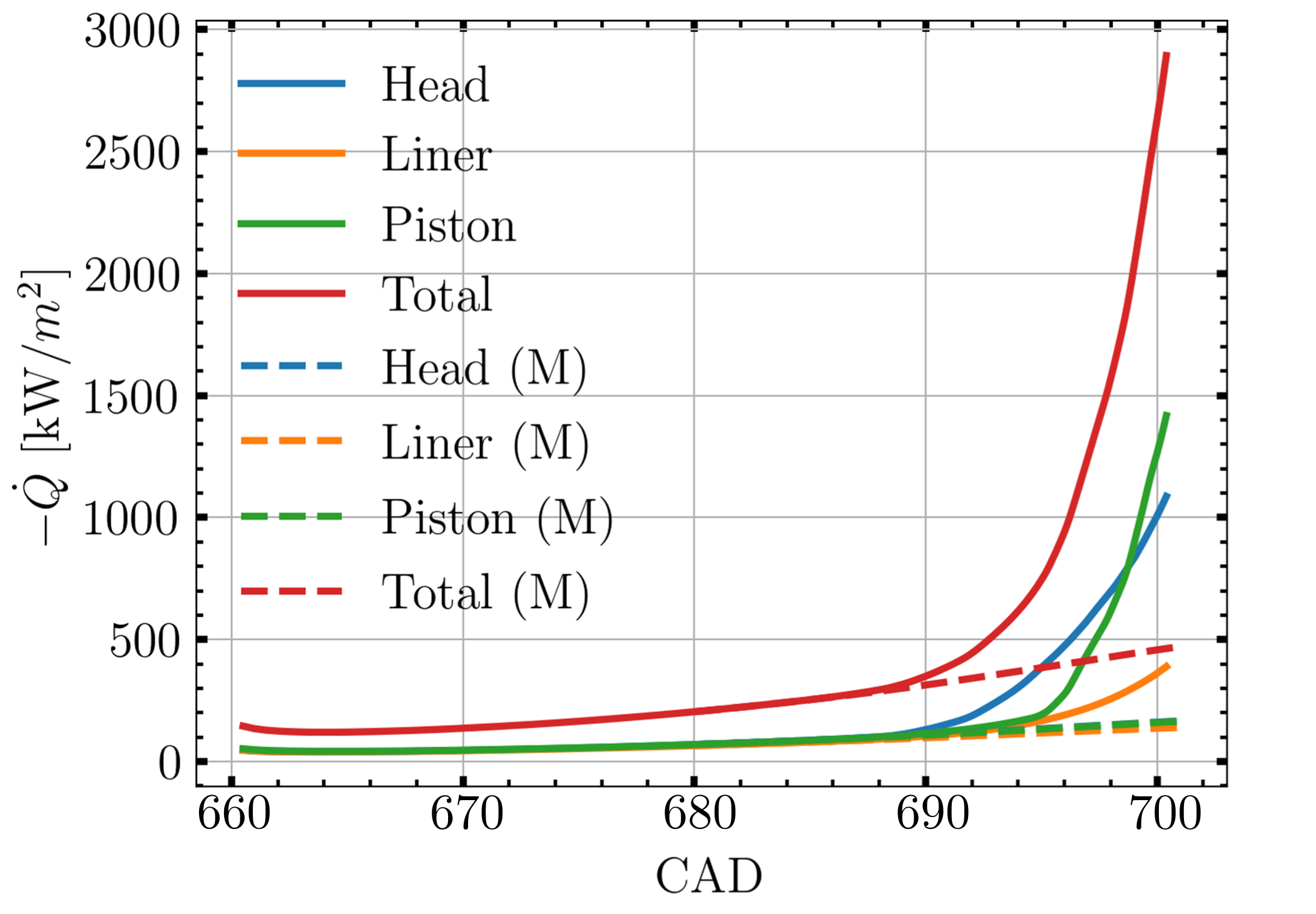}
\caption{Evolution of the component-wise surface-averaged heat flux.}
\label{fig:whf_component}
\end{figure}

In terms of the global metrics reported in Fig.~\ref{fig:whf_component}, the heat flux across different domain regions initially exhibit similar magnitudes and steadily increase before ignition as compression progresses. During this period, the cylinder head shows higher heat fluxes, as it is more exposed to impinging flow patterns induced by the tumble motion. After ignition, when the flame reaches the head near the spark plug and exhaust valves (at around 690 CAD), the wall heat flux increases abruptly, followed by a similar increase in the piston heat flux as the flame touches its surface at 695 CAD (Fig.~\ref{fig:whf_walls}d) and an exponential growth thereafter. The heat flux along the liner remains comparatively low, as the flame has not yet reached this area within the simulated CAD range. Overall, the reactive condition leads to heat flux values that are almost an order of magnitude higher than those observed in motored simulations where no reactions take place (solid vs. dashed lines in Fig.~\ref{fig:whf_component}).

\subsection{Flame quenching}

\begin{figure}[h!]
    \centering
    \hspace*{-1em}
    \begin{subfigure}{0.5\textwidth}
        \includegraphics[width=\textwidth]{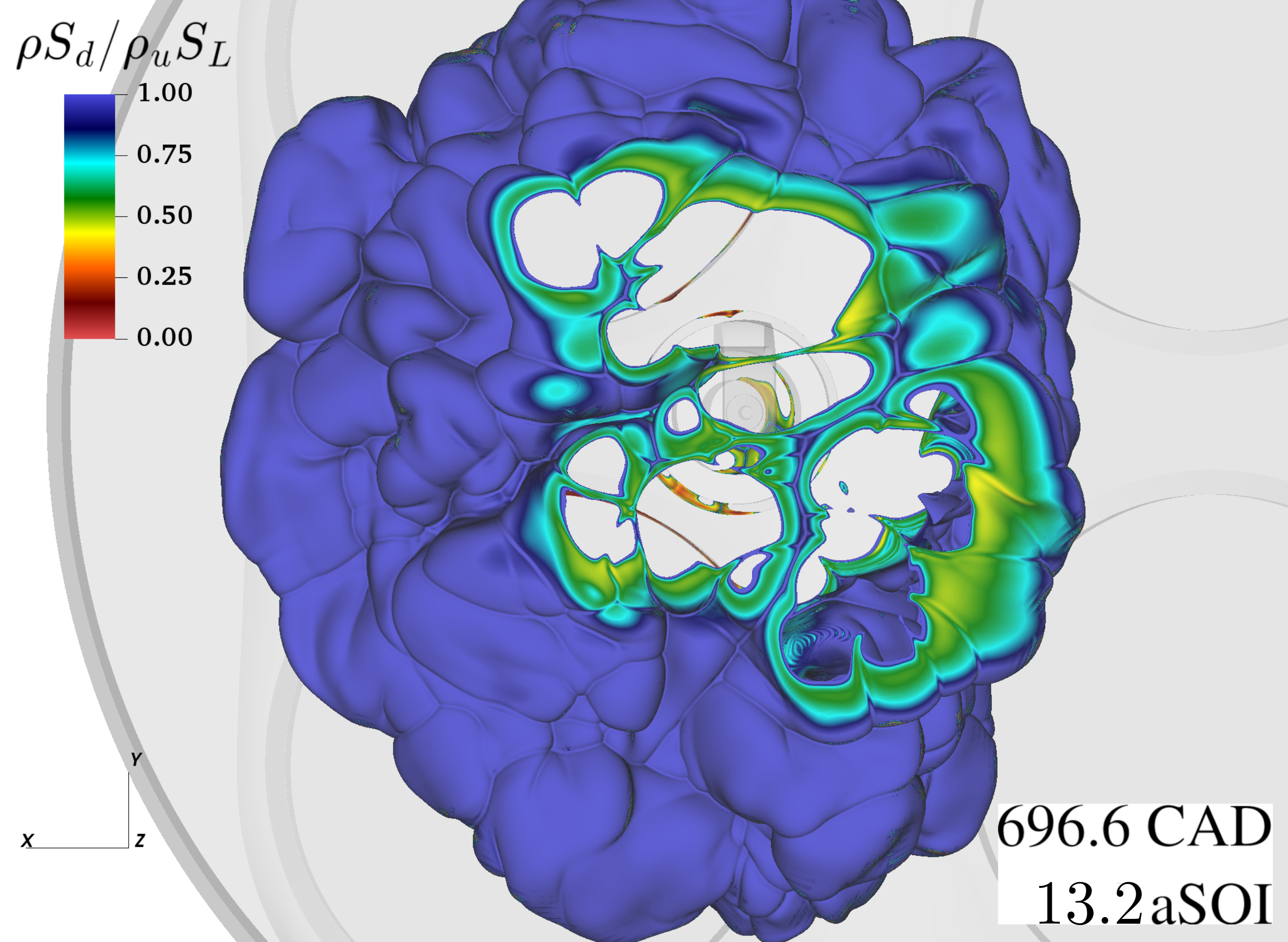}
        \caption{}
        \label{fig:quenching:a}
    \end{subfigure}
    \begin{subfigure}{0.5\textwidth}
        \includegraphics[width=\textwidth]{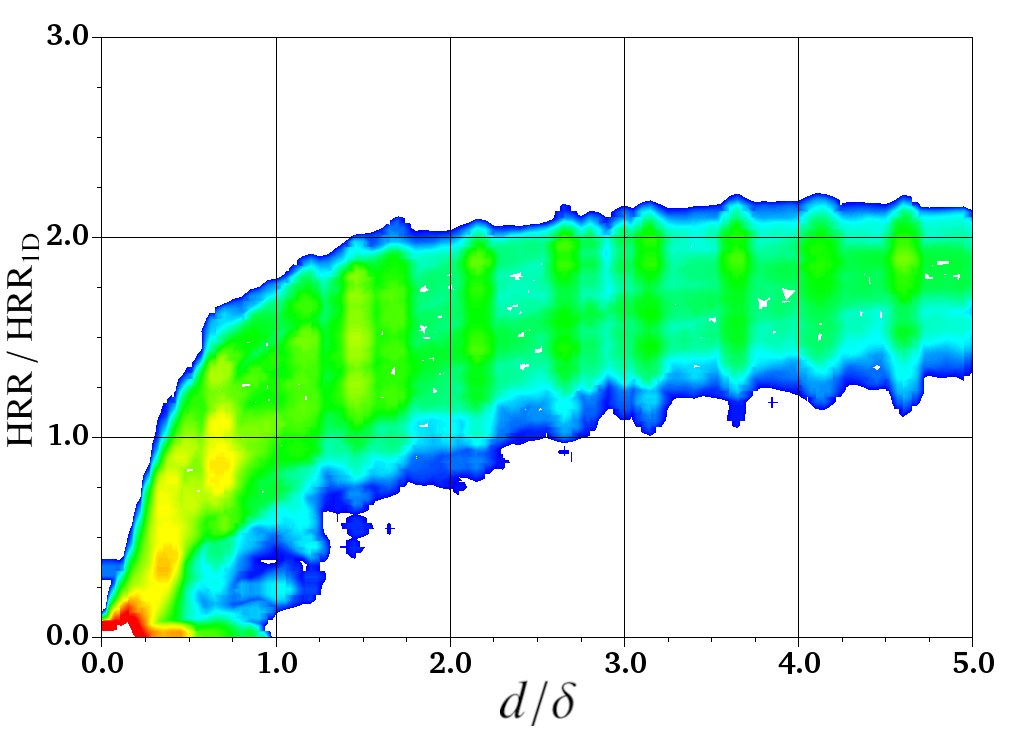}
        \caption{}
        \label{fig:quenching:b}
    \end{subfigure}
    \hspace*{-1em}
    \caption{(a) Bottom view of the flame surface colored by the density-weighted flame displacement speed and (b) joint pdf of the heat release rate (HRR) and flame-wall distance 13.2 CAD aSOI. All quantities are normalized with the corresponding 1D flame properties at the instantaneous thermodynamic conditions.}
    \label{fig:quenching}
\end{figure}

Figure~\ref{fig:quenching}(a) shows the flame surface colored by the density-weighted flame displacement speed at 13.2 CAD aSOI, viewed from below the piston along the positive $z$-axis. At this stage, the flame has impinged on the piston surface and is propagating parallel to it, indicating a side-wall quenching (SWQ) scenario, where the active part of the flame is located in the periphery of the contact region. 
At the center of the contact area with the piston the flame has remained close to the wall long enough to be fully quenched, as indicated by the disappearance of the $Y_{H_2}$=0.0048 iso-surface in this region. Indeed, this is the most appropriate definition of the flame front in wall-bounded configurations, as opposed for example to the use of a temperature iso-surface \cite{Jafargholi2018}. The color scale in the figure was adjusted to highlight the reduction in reactivity at the flame-wall interface in relation to its laminar equivalent, corresponding to the low $S_d$ branch noted in Figs.~\ref{fig:joint_pdfs}(j,k). 

A quantity of interest for the characterization of flame-wall interaction phenomena is the flame quenching distance $\delta_Q$, the shortest distance between the flame surface and the wall. It is usually reported in terms of a P\'{e}clet number $Pe=\delta_Q/\delta_l$, where $\delta_l=\alpha/S_L$ is the flame diffusion length scale with $\alpha$ the thermal diffusivity of the unburned mixture.
The joint pdf of the normalized heat release rate (HRR) and the flame-wall distance plotted in Fig.~\ref{fig:quenching}(b) shows that the flame reactivity starts to be affected by the wall at approximately one $\delta$, which corresponds to $Pe=8.2$. 
A more detailed analysis of flame-wall interaction phenomena, taking into account the spatio-temporally varying conditions as well as the effect of the local flow field in the vicinity of the wall, will be the subject of future work.

\section{Conclusions} \label{sec:conclusions}

In the present study, a direct numerical simulation (DNS) of hydrogen combustion in a real-size internal combustion engine (ICE) was performed and the results were analyzed focusing on ignition, flame propagation, and flame-wall interaction phenomena. The simulation utilized a state-of-the-art spectral element solver optimized for GPU architectures, resolving the relevant scales of turbulence and chemical reactions.

The results highlighted the complex interplay between the turbulent flow field, differential diffusion effects, and the time-varying thermodynamic conditions during the combustion process. The flame displacement speed exhibited a strong dependence on curvature, with positive curvatures enhancing reactivity due to differential diffusion effects, while negative curvatures were dominated by tangential diffusion contributions. The flame-wall interaction analysis identified different qualitative behavior during head-on (HOQ) and side-wall quenching (SWQ) scenarios. A first attempt to quantify the flame quenching distance is presented in a fully realistic engine configuration, showing higher P\'{e}clet values than previously reported for hydrogen/air mixtures, and a more elaborate analysis will be the subject of future work.

\section{Acknowledgements}

This project received funding from the European Union’s Horizon 2020 research and innovation program under the Center of Excellence in Combustion (CoEC) project, grant agreement No. 952181, the Forschungsvereinigung Verbrennungskraftmaschinen (FVV, Project No. 6015252) and the Swiss Federal Office of Energy (SFOE, Project No. SI/502670-01). The authors gratefully acknowledge the Gauss Centre for Supercomputing e.V. for providing computing time on the GCS Supercomputer JUWELS at J\"{u}lich Supercomputing Centre (JSC). This research used resources of the Argonne Leadership Computing Facility, which is a DOE Office of Science User Facility.

\section{References}

{\bibliographystyle{mcs13}
\setlength{\bibsep}{0.5mm}
\def\section*#1{}
\bibliography{H2TUDa}}

\end{document}